\documentclass{aa}
\usepackage[varg]{txfonts}
\usepackage[font=small,labelfont=bf]{caption}
\usepackage{graphicx} 
\usepackage[]{natbib, twoopt}
\usepackage[breaklinks=true]{hyperref} 
\usepackage[flushleft]{threeparttable}
\usepackage{adjustbox}
\usepackage{multirow}
\usepackage{amsmath,amssymb}
\usepackage{gensymb}
\usepackage{wasysym}
\usepackage{siunitx}
\usepackage{float}

\hypersetup{
      colorlinks=True,
      allcolors=blue
      }

\newcommand{\TableAone}{\hyperref[sec:data_availability]{A.1 }}
\newcommand{\TableBone}{\hyperref[sec:data_availability]{B.1 }}

\bibpunct{(}{)}{;}{a}{}{,}             
\makeatletter
  \newcommandtwoopt{\citeads}[3][][]{\href{http://adsabs.harvard.edu/abs/#3}%
    {\def\hyper@linkstart##1##2{}%
     \let\hyper@linkend\@empty\citealp[#1][#2]{#3}}}
  \newcommandtwoopt{\citepads}[3][][]{\href{http://adsabs.harvard.edu/abs/#3}%
    {\def\hyper@linkstart##1##2{}%
     \let\hyper@linkend\@empty\citep[#1][#2]{#3}}}
  \newcommandtwoopt{\citetads}[3][][]{\href{http://adsabs.harvard.edu/abs/#3}%
    {\def\hyper@linkstart##1##2{}%
     \let\hyper@linkend\@empty\citet[#1][#2]{#3}}}
  \newcommandtwoopt{\citeyearads}[3][][]%
    {\href{http://adsabs.harvard.edu/abs/#3}
    {\def\hyper@linkstart##1##2{}%
     \let\hyper@linkend\@empty\citeyear[#1][#2]{#3}}}
\makeatother

\begin{document}

  \title{A Song of Lines and Winds: \\ Tracing the Signatures of AGN Outflows in X-rays} 

  \titlerunning{}
  
  \subtitle{}

  \author{M.~Laurenti\inst{1,2,3,}\thanks{Corresponding author: \href{mailto:marco.laurenti@roma2.infn.it}{\color{black}\texttt{marco.laurenti@roma2.infn.it}}} \and F.~Tombesi\inst{1,2,3} \and P.~Condò\inst{1,4} \and M.~Gaspari\inst{5} \and F.~Nicastro\inst{3} \and E.~Torresi\inst{6} \and A.~Luminari\inst{3,7} \and E.~Piconcelli\inst{3} \and L.~Zappacosta\inst{3} \and K.~Fukumura\inst{8} \and G.~Lanzuisi\inst{6} \and R.~Serafinelli\inst{9,3} \and M.~Dadina\inst{6} \and M.~Cappi\inst{6} \and R.~Middei\inst{3,10} \and F.~ Arevalo Gonzalez\inst{1,3,4,11} \and F.~Di Salvo\inst{1,3}}
  
  \institute{Dipartimento di Fisica, Università degli Studi di Roma ``Tor Vergata'', via della Ricerca Scientifica 1, I-00133 Roma, Italy \and INFN $-$ Sezione di Roma ``Tor Vergata'', Via della Ricerca Scientifica 1, I-00133 Roma, Italy \and INAF $-$ Osservatorio Astronomico di Roma, via Frascati 33, I-00078 Monte Porzio Catone, Italy \and  Dipartimento di Fisica, ``Sapienza'' Università di Roma, Piazzale Aldo Moro 2, I-00185 Roma, Italy \and Department of Physics, Informatics and Mathematics, University of Modena and Reggio Emilia, 41125 Modena, Italy \and INAF $-$ Osservatorio di Astrofisica e Scienza dello Spazio di Bologna, Via Gobetti 101, I-40129 Bologna, Italy \and INAF $-$ Istituto di Astrofisica e Planetologia Spaziali, Via del Fosso del Caveliere 100, I-00133 Roma, Italy  \and Department of Physics and Astronomy, James Madison University, Harrisonburg, VA 22807, USA \and Instituto de Estudios Astrofísicos, Facultad de Ingeniería y Ciencias, Universidad Diego Portales, Av. Ejército Libertador 441, Santiago, Chile \and Space Science Data Center, Agenzia Spaziale Italiana, Via del Politecnico snc, 00133 Roma, Italy \and Jodrell Bank Centre for Astrophysics, School of Physics and Astronomy, University of Manchester, Manchester M13 9PL, UK }

  \date{}

  \abstract{Ultra-fast outflows (UFOs) are highly ionized, mildly relativistic winds seen in the X-ray spectra of active galactic nuclei (AGNs) and are thought to influence AGN feedback and galaxy evolution.}{In this work, we aim to investigate the UFO signatures by analyzing a large and diverse sample of detections from the literature.}{We compiled a sample of 122 solid (${>}\,2\sigma$ c.l.) UFO detections for a total of 57 AGNs, spanning broad ranges in redshift ($z\,{\lesssim}\,4$), bolometric luminosity ($10^{43}\,{\lesssim}\,L_\mathrm{bol}\,{\lesssim}\,10^{49}$ erg s$^{-1}$), black hole mass ($10^6\,{\lesssim}\, M_\mathrm{BH}/M_\odot\,{\lesssim}\,10^{10}$), and Eddington ratio ($-2.7\,{\lesssim}\,\log \lambda_\mathrm{Edd}\,{\lesssim}\,0.6$). We combined results from both phenomenological and photoionization modeling of the absorption features to characterize correlations among UFO parameters.}{We find evidence for a positive correlation between the line width $\sigma$, the equivalent width EW, and the outflow velocity $\upsilon_\mathrm{out}$, with the $\upsilon_\mathrm{out}-\sigma$ trend being comparatively weak. This suggests that the broadest absorption lines with the largest EW are signatures of the fastest UFOs.  
  We further demonstrate that the inferred velocity dispersion, often much larger than the uncertainty on the centroid velocity, should be accounted for in studies of wind energetics and scaling relations. We estimate lower limits on the launching radii of UFOs finding a minimum distance consistent with the innermost stable circular orbit (ISCO) of a weakly or non-rotating Schwarzschild black hole. This apparent truncation at smaller radii may reflect physical constraints, such as the presence of the X-ray corona or observational limitations due to line broadening. Additionally, variations in the line width to velocity ratio imply differences in wind geometry and kinematics. For the first time, we also explore the dependence of UFOs on AGN class. Differences in UFO properties between Seyferts and quasars \--- bridged by narrow-line Seyfert 1 galaxies \--- are likely driven by intrinsic parameters such as black hole mass and luminosity.}{The observed co-variation of $\upsilon_\mathrm{out}$ with both $\sigma$ and EW is consistent with clumpy, multi-component winds propagating through a thermally unstable multiphase medium within the chaotic cold accretion (CCA) cycle. These trends are not unique to a single acceleration mechanism: MHD and line-driven winds remain viable. High-resolution spectroscopy from missions like \emph{XRISM} and \emph{NewAthena} is necessary to fully resolve the structure, kinematics, and physical origin of UFOs.}


  \maketitle

\thinmuskip=3mu
\medmuskip=4mu
\thickmuskip=5mu

\nolinenumbers

\section{Introduction}\label{sec:intro}

Ultra-fast outflows (UFOs) are powerful winds launched in the central regions of active galactic nuclei (AGNs) and are usually detected as blue-shifted absorption lines in their X-ray spectra, likely attributable to highly ionized H- and He-like Fe K \citep[e.g.][]{Nardini2015, Tombesi2015, Luminari2018, Reeves2019, Serafinelli2019, Luminari2023, Reeves2023}.

Systematic studies \citep[e.g.][]{Tombesi2011, Gofford2013, Matzeu2023, Gianolli2024} have not only shown that the incidence of UFOs is relatively high, encompassing a fraction of the AGN population as large as $30{-}40\%$, but also indicated that the outflowing gas is highly ionized ($\log\xi \sim 3{-}6$ erg s$^{-1}$ cm), has a large column density ($N_\mathrm{H}\sim 10^{22-24}$ cm$^{-2}$) and is launched at mildly relativistic speeds ($\upsilon_\mathrm{out} \sim 0.03{-}0.5\,c$).
Moreover, evidence of UFOs has been reported in both type I and II AGNs, being either radio-quiet or radio-loud \citep[e.g.][]{Tombesi2010a, Tombesi2010b, Tombesi2014, Gofford2013, Braito2018, Middei2021, Mestici2024}.
Due to instrumental limitations, most of the UFOs have been detected in the local Universe ($z\lesssim0.1$) but some have been discovered in more distant (and typically lensed) AGNs \citep[e.g.][]{Chartas2002, Lanzuisi2012, Vignali2015, Dadina2018, Chartas2021}. 
UFOs carry such large kinetic power and momentum that they might play a key role in influencing galaxy evolution \citep[e.g.][]{DiMatteo2005, Hopkins2010, Gaspari2011, Wagner2013, Choi2014, Choi2015, Fiore2017, Gaspari2017, Costa2020}. Observational evidence includes cases where kpc-scale molecular outflows appear energetically consistent with being driven by inner UFOs \citep[e.g.][]{Feruglio2015, Tombesi2015} though much of the support remains indirect. UFOs are thus considered to be of pivotal importance in interpreting the tight correlations between the properties of the central supermassive black hole (SMBH) and its environment (see \citealt{Kormendy2013}, for a review), including the macro-scale hot halos \citep[e.g.][]{Gaspari2019}.  
In this broader feedback context, processes such as thermal instability (TI) and chaotic cold accretion (CCA) provide a natural framework for understanding how UFOs couple to their environment. TI arises when small perturbations in irradiated gas trigger runaway cooling, leading to the condensation of a multiphase medium in which cold, dense clumps coexist with a hot ambient plasma (e.g., \citealt{Field1965}; \citealt{McCourt2012}). In the CCA framework (e.g., \citealt{Gaspari2013, Gaspari2017}), such multiphase clouds condense out of the hot phase via nonlinear turbulent eddies and chaotically rain toward the SMBH. Their inelastic collisions with each other and the inner disk modulate the accretion rate and co-evolve with AGN-driven outflows in a self-regulated feedback loop (see \citealt{Gaspari2020} for a review). Similar TI–driven condensation has been identified in earlier simulations of radiatively heated accretion flows. In particular, \cite{Barai2012} and \cite{Moscibrodzka2013} showed that, under suitable X-ray illumination, hot gas can fragment into cold clumps and filaments that fall inward, producing a ``rain'' of material onto the black hole.
The CCA framework provides an analogous but independent picture, describing a condensation process driven primarily by turbulence within the hot halo and embedded in a self-regulated inflow–outflow cycle that connects large-scale cooling to SMBH feeding.
Hence, the physical mechanism \--- TI leading to multiphase rain \--- is shared across these models, although the dominant driving agent (central irradiation vs turbulent mixing) and spatial scale differ, making the two approaches complementary.

In recent years, UFOs have been observed with increasing frequency, yet the discourse on the physical mechanisms governing their launch and acceleration remains unresolved. Two leading scenarios are often considered: radiation-driven winds, which may arise either through continuum driving in super-Eddington sources or UV line-driving in sub-Eddington ones, and magnetically driven winds accelerated by the Lorentz force.
Ideally, different mechanisms should lead to unique imprints in the spectral shape of the absorption features, allowing us to distinguish between competing scenarios \citep[e.g.][]{Fukumura2022}. In practice, however, this is often hampered by the limited resolution of the CCD cameras onboard operating X-ray telescopes. As a result, a fine structure of multiple lines ascribed to a stratified absorber may be hidden beneath an apparently broad and smooth absorption profile.

The study of the profile and broadening of the absorption features attributable to UFOs is important not only for deciphering the launching mechanism but also for gaining insight into the velocity dispersion of the wind. Indeed, the outflow velocity $\upsilon_\mathrm{out}$ is often estimated either from the line centroid energy in phenomenological models or from the absorber redshift returned by photoionization codes such as XSTAR \citep[][]{Kallman2001}. However, these values usually quote only the statistical error on $\upsilon_\mathrm{out}$ and do not take into account the line width, which directly traces the velocity dispersion of the absorbing gas and offers complementary insight into the outflow kinematics.

The observed velocity dispersion lends itself to different interpretations. As discussed by \citet{Fukumura2019}, broadening has often been interpreted phenomenologically as turbulence intrinsic to the outflowing plasma, and it was noted that in simplified radiation-driven scenarios, the observed velocity gradients are not straightforward to reproduce. However, subsequent multi-dimensional simulations of radiation line-driven winds have shown that such flows are not strictly radial and can naturally generate broad absorption features via Doppler broadening (e.g. \citealt{Giustini2012}). In the same work, \citet{Fukumura2019} also proposed that, for a magnetohydrodynamic (MHD) wind, the line broadening may be due to transverse motion around a compact X-ray corona.  
Alternatively, the broadening can be understood under the assumption that the wind is launched from multiple zones located at different radii from the central SMBH, each contributing to the observed profile. This scenario is consistent with either a smooth wind launched from different areas of the accretion disk or a clumpy wind. The recent groundbreaking detection of five distinct velocity components in the broad and relativistic UFO in PDS 456 by \emph{XRISM} \citep[][]{PDS456} seems to support this latter view. Nonetheless, additional data from multiple sources are required to reach unequivocal conclusions.
Furthermore, we recall that knowledge of the line profile also has implications for the inferred energetics of the wind, such as the mass outflow rate \citep[e.g.][]{Gofford2015, Laurenti2021}, and its dispersion must be taken into account when studying possible correlations between the wind parameters. 

For this reason, in this paper, we present a sample of UFOs drawn from the literature to investigate, for the first time, the effects of their corresponding velocity dispersion. This dataset lays the groundwork for our aDvanced X-RAy modelinG Of quasar wiNds (DRAGON) project, which aims to employ the novel WINE photoionization code \citep[][]{Luminari2018, Luminari2024} on the largest currently available collection of X-ray quasar winds to derive the physical parameters of the outflows in a homogeneous and self-consistent way. Such an analysis is deferred to a forthcoming study, where we will explicitly address UFO spectral variability and line profile characterization, also leveraging new high-resolution data from XRISM as soon as they become publicly available. The description of the sample is provided in Sect.~\ref{sec:data}. The UFO properties are reviewed in Sect.~\ref{sec:ufo_properties}. The results are presented in Sect.~\ref{sec:results} and discussed in Sect.~\ref{sec:discussion}. Conclusions are left to Sect.~\ref{sec:conclusions}.
Throughout the paper, all errors are quoted at the $1\sigma$ confidence level, unless otherwise stated, while upper and lower limits are quoted at 90\% confidence level.

\section{Data and sample description}\label{sec:data}

The data presented in this paper are based on the results of the spectroscopic analyses carried out in a number of previous works that we describe in the following. \cite{Tombesi2010a} \--- T10a \--- embarked on a quest to discover UFO absorption features in a sample of 42 local ($z \lesssim 0.1$) radio-quiet AGNs observed by \textit{XMM-Newton}, comprising both type 1 and type 2 sources, and reported the presence of such powerful disk winds in ${\sim}\,35\%$ of the sample. T10a provided a phenomenological description of the absorption troughs in terms of Gaussian profiles. Later, \cite{Tombesi2011} thoroughly described the results of the XSTAR photoionization modeling of the Fe K absorption lines that were previously detected and presented in T10a.

Almost concurrently, \cite{Tombesi2010b} \--- T10b \--- conducted a systematic search for blueshifted Fe K absorption lines in the X-ray spectra of five broad-line radio galaxies observed by \textit{Suzaku}. The authors found that three of these sources were characterized by the typical spectral imprint of a UFO \--- namely, a blueshifted absorption feature \--- with high confidence. These absorption profiles were first addressed on a phenomenological basis as Gaussian lines and then modeled using the XSTAR photoionization code to provide a more physical and realistic view of the properties of the wind.

In the present paper, the AGN sample of \cite{Gofford2013} is also considered. Specifically, they used archival \textit{Suzaku} observations to perform a systematic blind search for blueshifted Fe K absorption features in a sample of 59 type 1-1.9 AGNs with a redshift of $z \lesssim 0.25$. According to the authors, the absorption attributable to \ion{Fe}{XXV} He$\alpha$ and/or \ion{Fe}{XXVI} Ly$\alpha$ is statistically found at the ${\gtrsim}\,2\sigma$ confidence level in approximately $40\%$ of the sample.

\cite{Tombesi2014} followed in the footsteps of T10b and analyzed a larger sample of 26 RL AGNs ($z \lesssim 0.2$) to probe the incidence of UFO features in this AGN class by leveraging data collected by \textit{Suzaku} and \textit{XMM-Newton}. They detected UFOs in $>27\%$ of the sources and reported the results derived from both the phenomenological and physical photoionization modeling of the wind.

A detailed search for UFOs in high-$z$ AGNs was carried out by \cite{Chartas2021} \--- C21 \--- who considered a sample of 14 quasars (QSOs) with a redshift of $1<z<4$, most of which are gravitationally lensed. C21 aimed to provide a physical description of the wind properties by means of photoionization modeling with XSTAR. C21 analyzed only six of the 14 QSOs in their sample, while the results for the remaining objects were based on previous independent studies. In order to collect information on the phenomenological description of the UFO absorption features in these QSOs, we relied on these studies. Specifically, for APM 08279+5255 we used the results from \citet{Chartas2002, Chartas2009}, for HS 1700+6416 from \cite{Lanzuisi2012}, for MG J0414+0534 from \cite{Dadina2018}, for PG 1115+080 from \citet{Chartas2003, Chartas2007}, for PID352 from \cite{Vignali2015}, and for HS 0810+2554 from \cite{Chartas2016}.

\cite{Matzeu2023} recently performed a systematic search for UFO features in a sample of 22 luminous ($L_\mathrm{bol}\sim10^{45-46}$ erg s$^{-1}$) AGNs at intermediate redshift ($0.1<z<0.4$). Blueshifted absorption lines due to highly ionized iron were significantly detected at a $\geq95\%$ confidence level in 7 out of the 22 AGNs, i.e., in $\sim30\%$ of their sample that we include in this work. 

We also consider a collection of highly significant individual UFO detections reported in independent studies. Specifically, for IRAS 18325--5926 we used the results in \cite{Iwasawa2016}, for IRAS F11119+3257 we considered the analysis of \cite{Tombesi2015}, for PG 1001+291 we collected information from \cite{Nardini2019}, for PG 1448+273 we used the results of \cite{Laurenti2021} and \cite{Reeves2023, Reeves2024}, for WKK 4438 we retrieved the results from \cite{Jiang2018}, for NGC 2992 we considered the time-resolved analysis carried out by \cite{Luminari2023}, for IZw1 we relied on the work of \cite{Reeves2019}, for IRAS 05189--2524 we used the results from \cite{Smith2019}, and for IRAS 13224--3809 from \cite{Chartas2018}, while for PDS 456 we also considered the results in \cite{Nardini2015}.

\begin{figure}[t]
    \centering
    \includegraphics[width=\columnwidth]{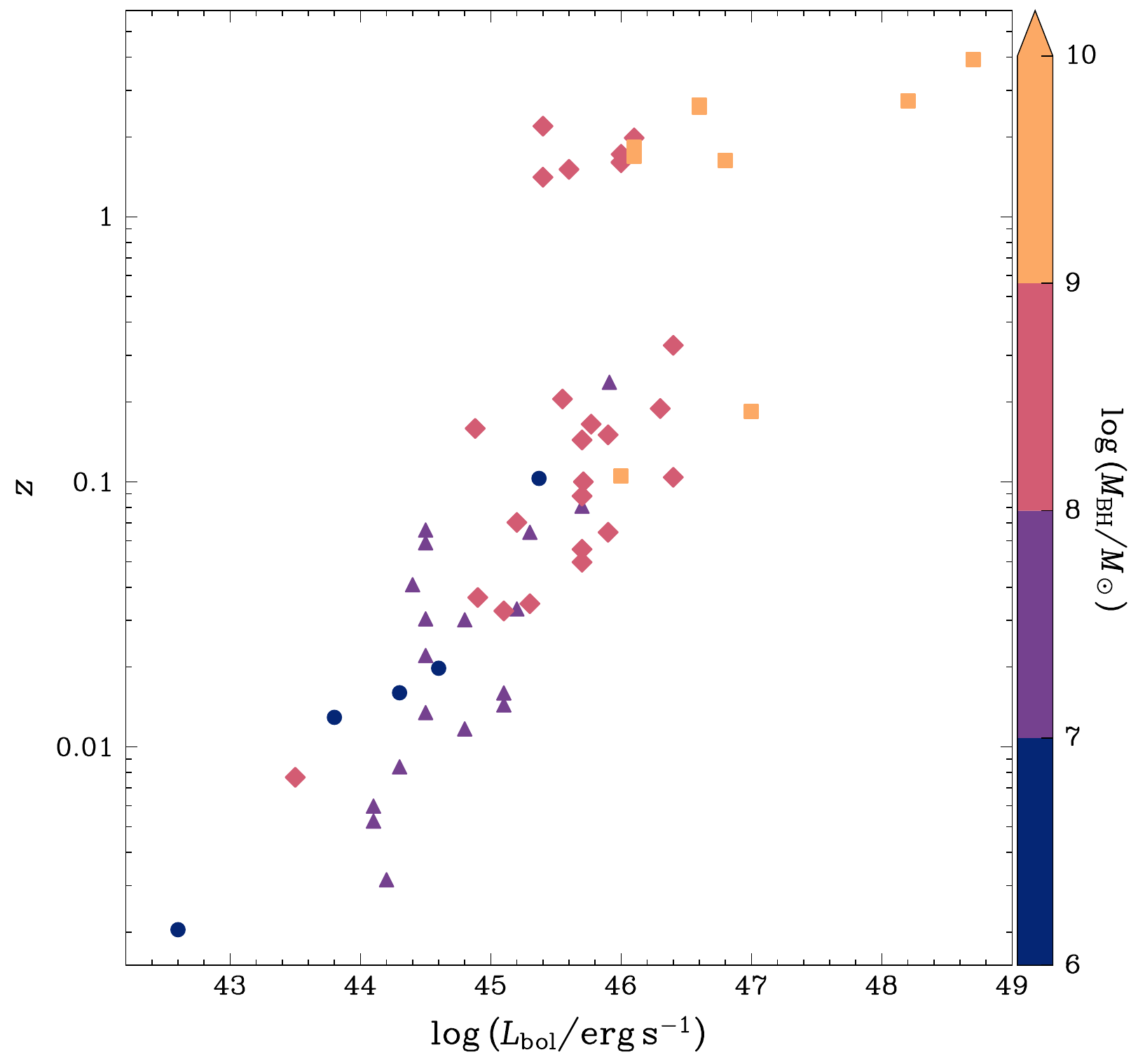}
    \caption{Distribution of our AGN sample in the $z-L_\mathrm{bol}$ plane. The color bar helps identifying the black hole mass distribution of these sources, split into four $M_\mathrm{BH}$ intervals described by different markers.  }
    \label{fig:L_z}
\end{figure}

To summarize, we merged all the results from the aforementioned studies and obtained a parent sample of 66 AGNs, resulting in a total of 158 solid detections of blueshifted iron K absorption troughs. We then selected only those winds whose speed was well constrained and consistent with the canonical definition of a UFO -- that is, $\upsilon_\mathrm{UFO}\geq10\,000$ km s$^{-1}$. 
Finally, we gathered a sample of 122 UFO detections, for a total of 57 AGNs spread across wide intervals in redshift ($z\lesssim4$), bolometric luminosity ($10^{43} \lesssim L_\mathrm{bol} \lesssim 10^{49}$ erg s$^{-1}$), black hole mass ($10^6 \lesssim M_\mathrm{BH}/M_\odot \lesssim 10^{10}$) and Eddington ratio ($-2.7 \lesssim \log{\lambda_\mathrm{Edd}} \lesssim 0.6$). A detailed description of our sample sources is provided in Table \TableAone along with their main properties. Figure \ref{fig:L_z} shows the distribution of our sample sources in the $z-L_\mathrm{bol}$ plane, where the AGNs are divided into four different $M_\mathrm{BH}$ intervals. We also highlight that $\gtrsim40\%$ of the AGNs in our sample possess multi-epoch UFO detections. In the following, for these sources in which the UFO is persistent, we will consider the main wind parameters as those corresponding to their median values across all the available epochs. We also checked that the results presented in the following sections remain unchanged when adopting mean values.

\section{UFOs properties}\label{sec:ufo_properties}

\renewcommand*{\ttdefault}{qcr}

For each UFO detection, we collected results from both the phenomenological and physical modeling of the wind, as reported in the works summarized in Sect.~\ref{sec:data}. In the former, the absorption feature is modeled with an inverted Gaussian profile, and we retrieved the rest-frame energy of the line centroid $E_\mathrm{rest}$, its width $\sigma$, which reflects the velocity dispersion usually attributed to turbulence or velocity shear in the outflow, and the equivalent width EW, which measures the total absorption strength and depends on the column density and the ionization state of the absorbing gas. From the photoionization modeling with XSTAR, we obtained the values of the UFO velocity $\upsilon_\mathrm{out}$, its column density $\log{N_\mathrm{H}}$ and the ionization parameter $\log\xi$.
While the Doppler shift of the Gaussian absorption line centroid provides a first-order estimate of $\upsilon_\mathrm{out}$, XSTAR simultaneously accounts for multiple ionic transitions, thereby reducing line-identification degeneracies (e.g. Fe XXV He$\alpha$ vs. Fe XXVI Ly$\alpha$) and providing a more physically motivated velocity measurement that is generally consistent with velocities derived from the line centroid \citep[e.g.][]{Tombesi2010a,Tombesi2010b,Tombesi2011}. For this reason, we adopt $\upsilon_\mathrm{out}$ from XSTAR fits as our reference value throughout this work.
The UFO properties are stored in Table \TableBone. 

\begin{figure}[t]
    \centering
    \includegraphics[width=\columnwidth]{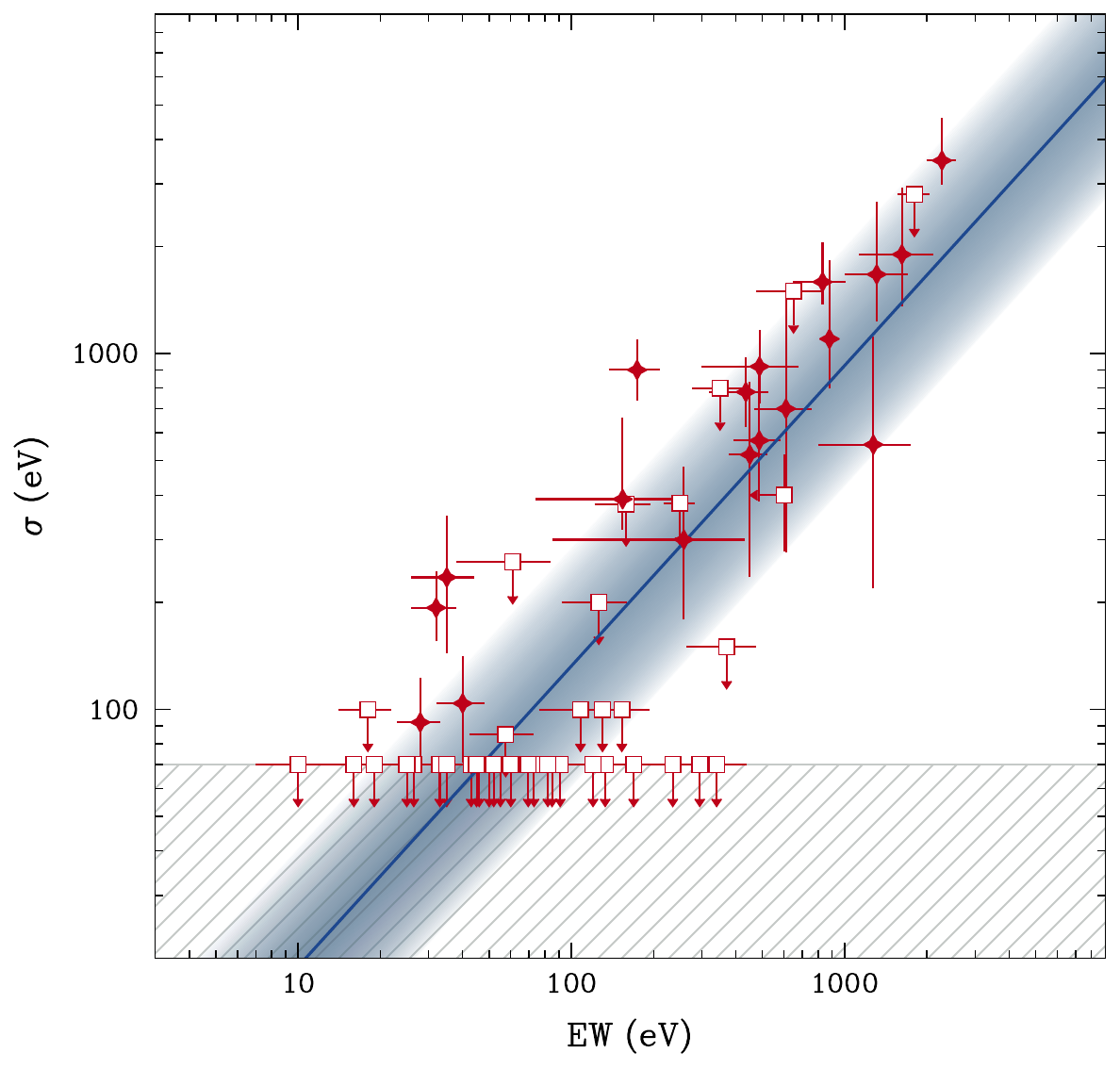}
    \caption{Distribution of UFOs for each AGN in our sample in the $\sigma - \mathrm{EW}$ plane. Solid blue line and shaded areas describe the best fit relation and its $1\sigma$ spread, respectively. Hatched region marks the average threshold on the line width for the current instrumental CCD resolution in the Fe K band. White squares are the UFOs with upper limits in $\sigma$ or EW.}
    \label{fig:sigma_vs_ew}
\end{figure}

All but one AGN in our sample have well-constrained measurements of the line's EW. Instead, the line width $\sigma$ is rarely well constrained, as this occurs only for 19 out of a total of 57 AGNs (${\sim}\,33\%$). The distribution of our sample sources in the $\sigma-\mathrm{EW}$ plane is shown in Fig.~\ref{fig:sigma_vs_ew}. The plane is largely populated by upper limits, especially in $\sigma$, which must be taken into account when performing the correlation analysis. We specify that, in doing so, we impose a homogeneous upper limit of $\sigma<70$ eV on those lines with unresolved width \--- whose fraction amounts to $\sim66\%$ of the whole UFOs with unconstrained $\sigma$ \--- which roughly corresponds to the limiting instrumental resolution of present-day CCD cameras. First, to assess the correlation strength and its statistical significance, we use the R package \texttt{cenken}, which implements a generalized version of Kendall’s $\tau$ correlation test, incorporating upper limits in both the predictor and response variables \citep{Helsel2005}. For the objects in Fig.~\ref{fig:sigma_vs_ew}, we find a Kendall correlation coefficient of $\tau\simeq0.33$ for a corresponding null probability of $p({>}|\tau|) \simeq 10^{-4}$. 

The best-fit line is calculated employing a bootstrapping procedure, carried out by resampling $N=100\,000$ times from the original distribution in Fig.~\ref{fig:sigma_vs_ew}, while accounting for the uncertainties in both variables as described below. In each iteration, the points uncensored in both $\sigma$ and EW (red diamonds) are updated by sampling from a normal distribution whose mean and variance correspond to their original values and associated uncertainties, respectively.
We treat upper limits (white squares) similarly to uncensored data, but with appropriate constraints.
In fact, we obtain an estimate of their value by drawing from a uniform distribution $\mathcal{U}(m,\sigma_\mathrm{ul})$ or $\mathcal{U}(n,\mathrm{EW}_\mathrm{ul})$ bounded between a custom nonzero value and the reported upper limit $\sigma_\mathrm{ul}$ or $\mathrm{EW}_\mathrm{ul}$. Specifically, we choose $m=5$ eV and $n=10$ eV for the upper limits in $\sigma$ and EW, respectively. To determine their associated uncertainty, we multiply their nominal values $\sigma_\mathrm{ul}$ and $\mathrm{EW}_\mathrm{ul}$ by the median relative error in either $\sigma$ or EW calculated from the uncensored data points. 
We tested that the following results are not sensitive to variations of up to a factor of ${\times}5$ in $m$ and/or $n$. 

\begin{figure}[t]
    \centering
    \includegraphics[width=\columnwidth]{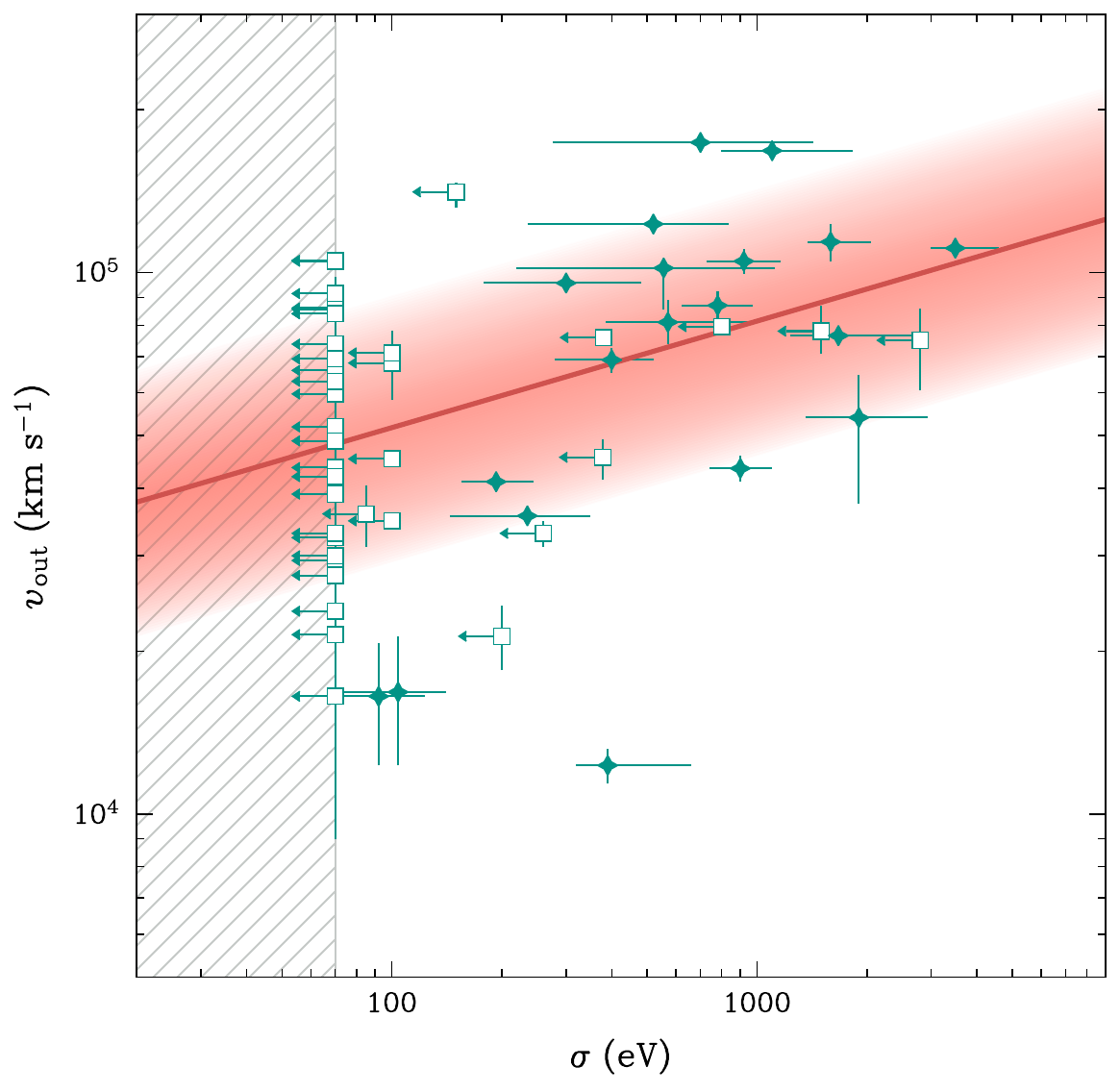}
    \caption{UFOs distribution in the $\upsilon_\mathrm{out} - \sigma$ plane for our sample sources. Solid red line and shaded areas describe the best fit relation and its $1\sigma$ spread, respectively. Hatched region marks the average threshold on the line width for the current instrumental CCD resolution in the Fe K band. White squares are as in Fig.~\ref{fig:sigma_vs_ew}.}
    \label{fig:v_vs_sigma}
\end{figure}

After each sampling, the best-fit line is determined using \texttt{linmix}\renewcommand*{\ttdefault}{cmr}\footnote{\url{https://github.com/jmeyers314/linmix}.}, \renewcommand*{\ttdefault}{qcr}which is the Python adaptation of the \texttt{LINMIX\_ERR} IDL package detailed by \citet{Kelly2007}. This approach utilizes a hierarchical Bayesian framework for executing linear regression, conscientiously considering errors in both dependent and independent variables. When compared to other statistical techniques, \texttt{linmix} exhibits greater robustness and provides an unbiased estimation even with minimal sample sizes and significant uncertainties \citep[e.g.][]{Sereno2016}. Assigning error bars to censored variables allows us to harness the complete capabilities of \texttt{linmix}, effectively overcoming its limitations. This approach is especially valuable because, as noted in the documentation, the maximum likelihood estimate is insufficient when handling censored datasets and is incapable of dealing with doubly censored variables.

The above procedure is repeated $N=100\,000$ times, and we collect all the slope and intercept values calculated by \texttt{linmix}, along with their corresponding $1\sigma$ errors, as well as the intrinsic scatter of the regression. Finally, the best-fit line is obtained by considering the median values of such distributions.
Specifically, in this case, we find:
\begin{equation}\label{eq:sigma_vs_ew}
    \log\sigma =  (0.8 \pm 0.1) + (0.4 \pm 0.2)\,\log\mathrm{EW}\,,
\end{equation}

\noindent which is characterized by a $1\sigma$ spread of $\sim0.3$ dex. This result suggests that the absorption features with the largest EW are also the broadest.

We adopt the same strategy described above to study the distribution of our sample sources in the $\upsilon_\mathrm{out} - \sigma$ plane shown in Figure \ref{fig:v_vs_sigma}. In this case, we find a Kendall correlation coefficient of $\tau \simeq 0.20$ for a corresponding $p({>}|\tau|) \simeq 0.036$. The best fit relation is characterized by a $1\sigma$ spread of about $\sim0.25$ dex, and can be expressed as follows:
\begin{equation}\label{eq:v_vs_sigma}
    \log\upsilon_\mathrm{out} =  (4.34 \pm 0.15) + (0.19 \pm 0.07)\,\log\sigma\,.
\end{equation}

\noindent Although this correlation is weak and marginally significant, it suggests a positive trend between $\upsilon_\mathrm{out}$ and $\sigma$.

\begin{figure}[t]
    \centering
    \includegraphics[width=\columnwidth]{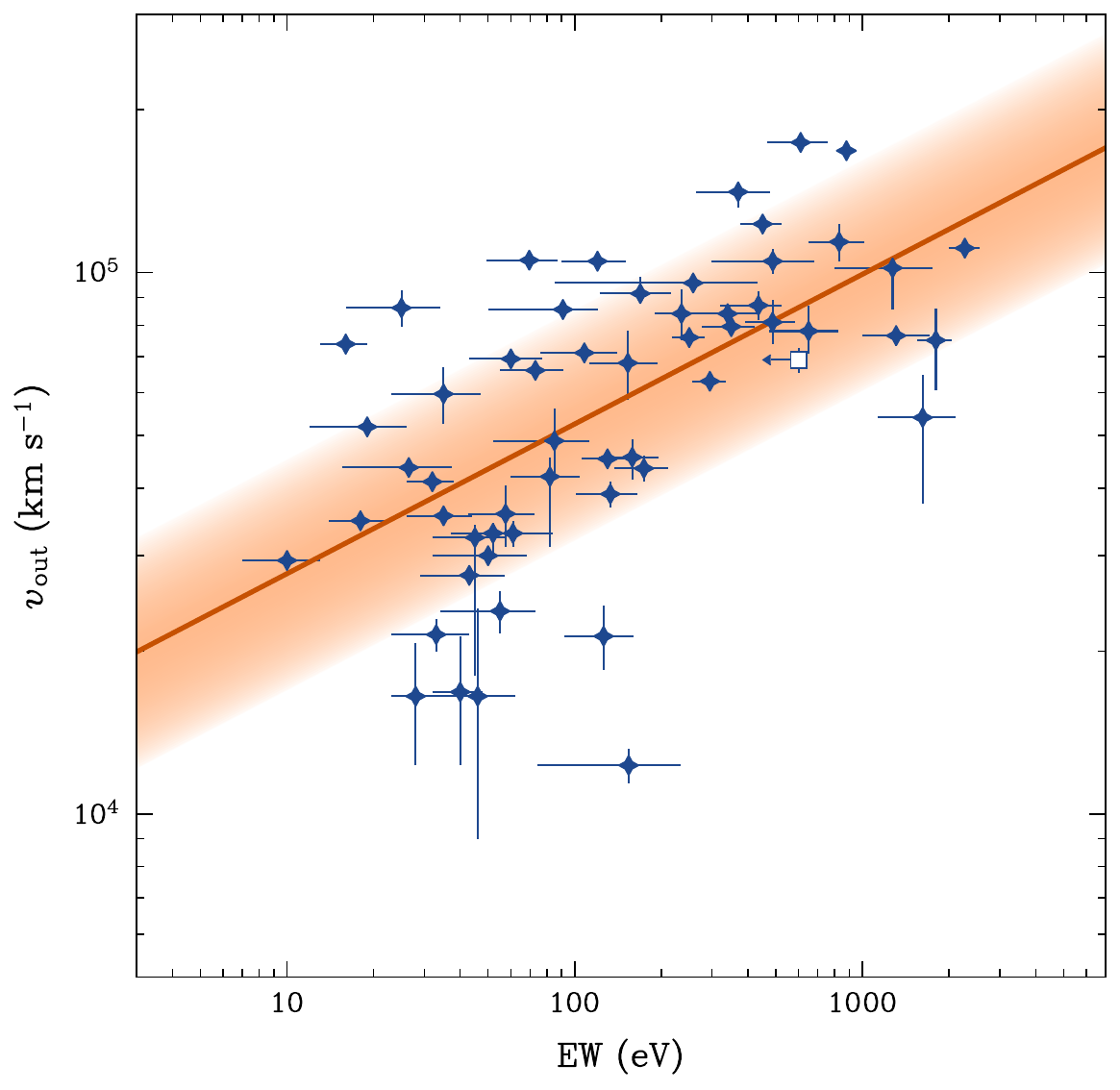}
    \caption{Distribution of UFOs in the $\upsilon_\mathrm{out} - \mathrm{EW}$ plane for our sample sources. Solid orange line and shaded areas describe the best fit relation and its $1\sigma$ spread, respectively.}
    \label{fig:v_vs_ew}
\end{figure}

Moreover, we find a highly significant correlation between the outflow velocity $\upsilon_\mathrm{out}$ and EW. The Kendall correlation coefficient amounts to $\tau \simeq 0.44$ while the corresponding null probability is approximately $p({>}|\tau|) \simeq 1.7 \times 10^{-6}$.
Figure \ref{fig:v_vs_ew} shows the distribution of our sources in the $\upsilon_\mathrm{out} - \mathrm{EW}$ plane.
In this case, we obtain a best fit relation of the form:
\begin{equation}\label{eq:v_vs_sigma}
    \log\upsilon_\mathrm{out} =  (4.2 \pm 0.1) + (0.28 \pm 0.05)\,\log\mathrm{EW}\,,
\end{equation}

\noindent whose $1\sigma$ spread is about $\sim0.21$ dex. According to Eq.~\ref{eq:v_vs_sigma}, the fastest UFOs are likely characterized by the most intense absorption features in the X-ray spectrum.

\begin{figure*}[t]
    \centering
    \includegraphics[width=\textwidth]{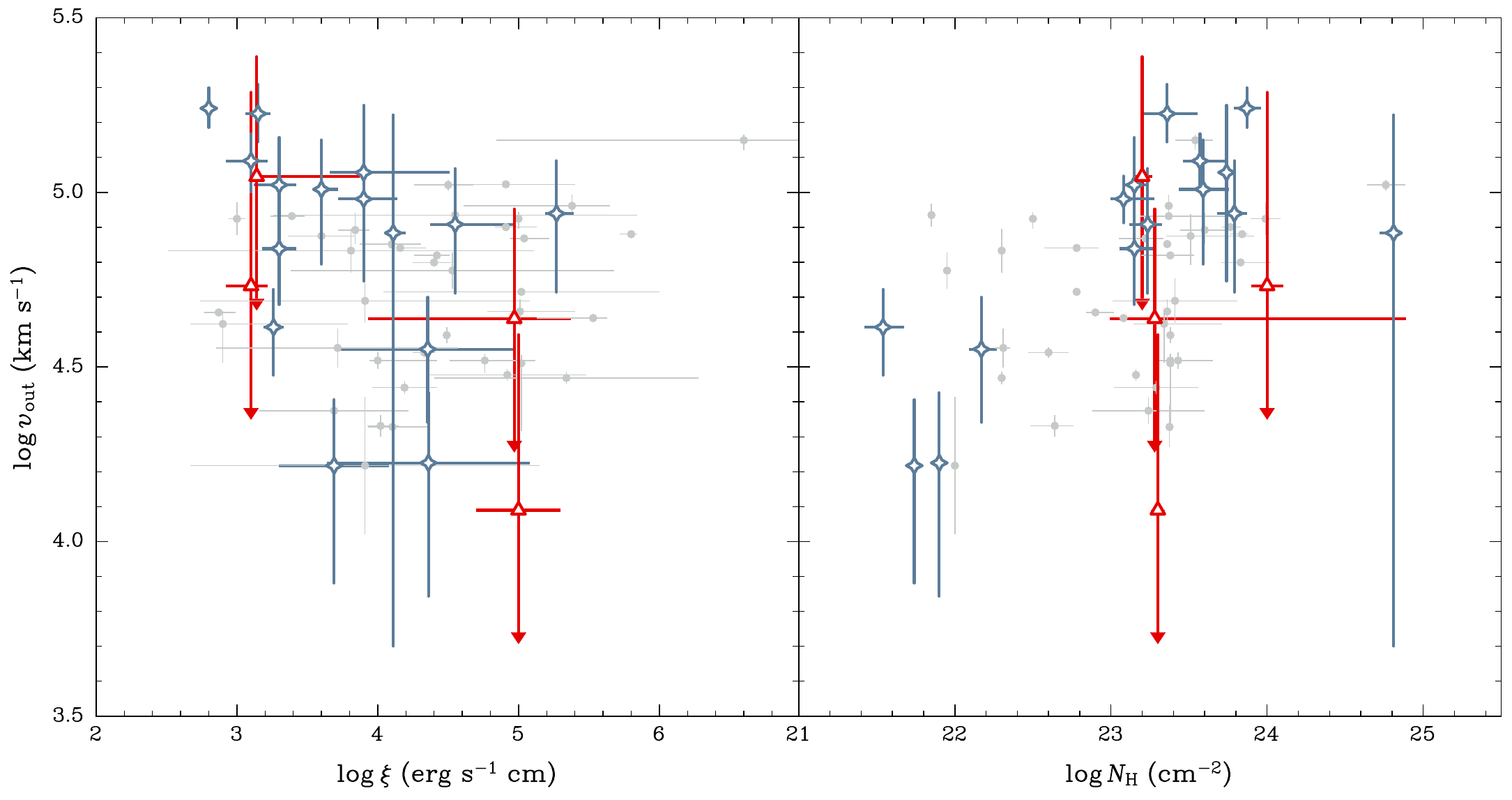}
    \caption{Distribution of UFO parameters in the $\upsilon_\mathrm{out} - \log{\xi}$ (left panel) and $\upsilon_\mathrm{out} - \log{N_\mathrm{H}}$ (right panel) planes for our sample. The vertical bars indicate the overall velocity dispersion of the wind in terms of the FWHM of the absorption line. Those AGNs whose UFO velocity is consistent with zero within the global dispersion, namely 3C 111, Q2237+0305, 4C+74.26 and PG 1115+1080 are shown in red. Grey dots represent the UFOs with unconstrained values of $\sigma$.}
    \label{fig:lgxi_lgnh_vs_v_fwhm}
\end{figure*}

\section{Results}\label{sec:results}

The $\upsilon_\mathrm{out}$ measurements and their corresponding uncertainties are drawn directly from the best fit returned by either phenomenological or physical modeling of the wind absorption in the rest-frame Fe K band. This implies that the actual width of the line profile and, thus, the velocity dispersion are usually not taken into account when studying possible correlations involving $\upsilon_\mathrm{out}$ and other relevant quantities. For this reason, it would be highly desirable to obtain a more realistic description of the characteristic range of $\upsilon_\mathrm{out}$ values that the given UFO can possibly achieve.

\subsection{Velocity dispersion}

When we phenomenologically model the UFO absorption profile with an inverted Gaussian, we obtain an estimate of its energy $E$ and width $\sigma$. If we want to consider these quantities in the outflow rest frame, we should take into account a cumulative correction factor of $1+z_\mathrm{tot}=(1 + z_\mathrm{c})(1 + z_\mathrm{out})$, where $z_\mathrm{c}$ and $z_\mathrm{tot}$ represent the cosmological redshift and the observed absorber redshift, respectively. The intrinsic absorber redshift $z_\mathrm{out}$, instead, can be directly computed from the relativistic Doppler formula $z_\mathrm{out}=\sqrt{(1-\beta)/(1+\beta)} -1$, where $\beta=\upsilon_\mathrm{out}/c$ is required to be positive for an outflow. Specifically, the line width in the outflow rest frame is $\sigma_\mathrm{r} = (1 + z_\mathrm{tot})\,\sigma$.

In order to relate the line width (the spread of energies around the centroid) to the velocity dispersion of the outflow, we can use the same relation as above and express the energy $E_\mathrm{r}$ of the centroid in the rest frame of the outflow in terms of the observed energy $E$ as:

\begin{equation}\label{eq:E_r}
    E_\mathrm{r} = E\,(1+z_\mathrm{tot}) = E\,(1+z_\mathrm{c})\,\sqrt{\frac{1-\beta}{1+\beta}}\,.
\end{equation}

\noindent Next, we can differentiate Eq.~\ref{eq:E_r} with respect to $\beta$ and then approximate the result for small finite changes.
Indeed, for an outflow moving at velocity $\upsilon_\mathrm{out}$, the fractional variation of the rest frame energy $E_\mathrm{r}$ of the line centroid with respect to $\beta$ is:
\begin{equation}\label{eq:fractional_var}
    \frac{\Delta{E_\mathrm{r}}}{E_\mathrm{r}} = \frac{\Delta\beta}{1-\beta^2} = \frac{\Delta\upsilon}{c}\,\frac{1}{1-\beta^2}
\end{equation}

\noindent By interpreting the line width in terms of energy and velocity as $\Delta{E_\mathrm{r}}=\sigma_\mathrm{r}$ and $\Delta\upsilon=\sigma_\mathrm{\upsilon}$, respectively, we can use Eq.~\ref{eq:fractional_var} to derive an estimate of the line width in units of km s$^{-1}$ as:
\begin{equation}\label{eq:sigma_v}
    \sigma_\mathrm{\upsilon} = \frac{\sigma_\mathrm{r}}{E_\mathrm{r}}\,(1-\beta^2)\,c\,,
\end{equation}

\noindent which is always smaller than the speed of light under the reasonable assumption that $\sigma_\mathrm{r}\,{<}\,E_\mathrm{r}$. Finally, since we are dealing with Gaussian profiles, we can obtain an estimate of the full width at half maximum (FWHM) of the absorption line according to its standard definition, exploiting Eq.~\ref{eq:sigma_v}:
\begin{equation}\label{eq:fwhm}
    \mathrm{FWHM} = 2\sqrt{2\ln{2}}\,\sigma_\upsilon \approx 2.355\, \frac{\sigma_\mathrm{r}}{E_\mathrm{r}}\,(1-\beta^2)\,c\,.
\end{equation}

\noindent Using the FWHM measurement, one can finally determine the larger range of velocities that the outflow may possibly reach. In fact, such a range would be represented by a spread of $\pm\,(\mathrm{FWHM} / 2)$ around the best-fit estimate of $\upsilon_\mathrm{out}$.

The two panels of Figure \ref{fig:lgxi_lgnh_vs_v_fwhm} show the distribution of our sample sources in the $\upsilon_\mathrm{out} - \log{\xi}$ and $\upsilon_\mathrm{out} - \log{N_\mathrm{H}}$ planes when we take into account the global velocity dispersion of the wind. One can notice that each UFO is characterized by a wide range of possible velocities. In the case of 3C 111, Q2237+0305, 4C+74.26 and PG 1115+1080, the lower end of the velocity dispersion is even consistent with zero, suggesting a range of velocities starting from rest up to the highest blue-shifted value.  

\begin{figure}[t]
    \centering
    \includegraphics[width=\columnwidth]{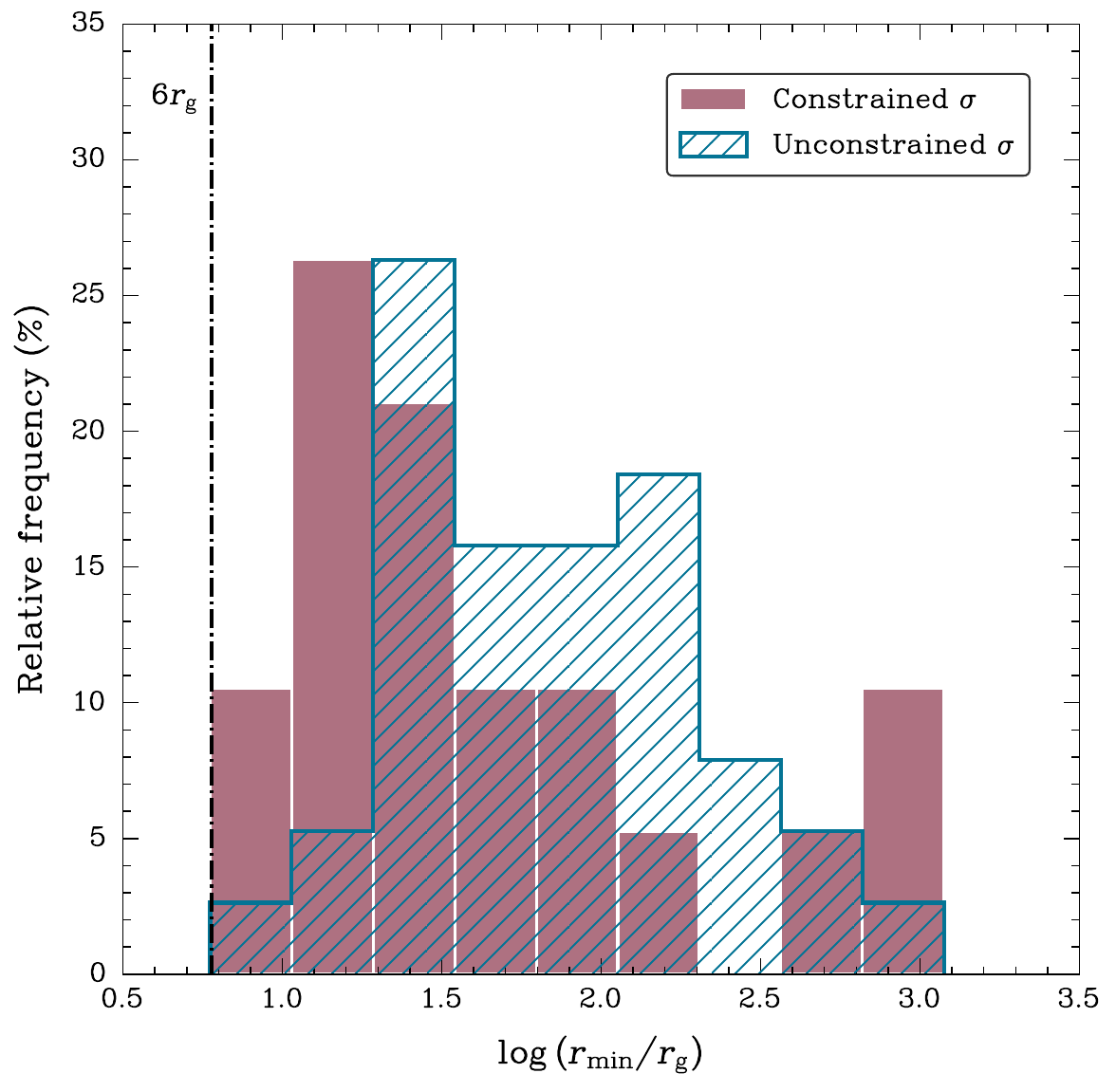}
    \caption{Relative frequency histogram showing the distribution of the lower limits $r_\mathrm{min}$ for the launching radius in units of $r_\mathrm{g}$ for the two subsamples with constrained (in red) and unconstrained line width (in blue), respectively. Dash-dotted black line indicates the location of the ISCO around a non-rotating Schwartzschild black hole.}
    \label{fig:rmin}
\end{figure}

\subsection{Wind geometry and kinematics}\label{sec:radius}

The resolution of present day CCD cameras poses a severe limitation on our ability to identify the exact location of the UFOs. A well-established practice \citep[e.g.][]{Tombesi2013, Gofford2015, Chartas2021} to obtain a rough estimate of the lower bound of the distance of the absorber from the central SMBH, $r_\mathrm{min},$ amounts to finding the radius at which the gas reaches the local escape velocity:
\begin{equation}
    r_\mathrm{min} = \frac{2 G M_\mathrm{BH}}{\upsilon^2_\mathrm{out}}\,,
\end{equation}

\noindent where $\upsilon_\mathrm{out}$ is the best fit value for the outflow velocity from XSTAR.
The distribution of the lower limits $r_\mathrm{min}$ for the location of the absorbers in units of gravitational radii $r_\mathrm{g}=GM_\mathrm{BH}/c^2$ is shown in Figure \ref{fig:rmin}.  
Interestingly, no sources in our sample show values below $\sim 6\,r_\mathrm{g}$, producing an apparent truncation toward smaller radii. This radius corresponds to the theoretical location of the innermost stable circular orbit (ISCO) around a non-rotating Schwarzschild black hole. Whether this truncation is physical or simply reflects small-number statistics remains uncertain; a formal test of significance would require assumptions about the expected underlying distribution, and we defer such an analysis to future work with larger samples. Nonetheless, the apparent truncation near the ISCO is consistently observed in both subsamples of UFOs shown in Fig.~\ref{fig:rmin}, regardless of whether the line width is constrained or unconstrained.
According to a Kolmogorov-Smirnov (KS) test, the null hypothesis cannot be rejected ($p{-}\mathrm{value} \simeq 0.06$). Diverse methodologies, including the Mann-Whitney U test, yield identical conclusions. Operatively, we perform these tests by employing the \texttt{ks\_2samp} and \texttt{mannwhitneyu} functions in SciPy. This result suggests that all UFOs may originate from the same range of distances from the SMBH.

Moreover, the ratio between the FWHM and $\upsilon_\mathrm{out}$ provides a measure of the velocity-space extent of the absorber relative to its bulk motion. If broadening is dominated by turbulence, this ratio reflects the relative strength of turbulent velocities compared to the outflow velocity. If broadening arises instead from multiple absorbing components, the same ratio quantifies the spread in velocity among the components relative to the systemic outflow speed. In both cases, FWHM/$\upsilon_\mathrm{out}$ can thus provide useful information about the kinematic and structural properties of the outflowing gas.

\begin{figure}[t]
    \centering
    \includegraphics[width=\columnwidth]{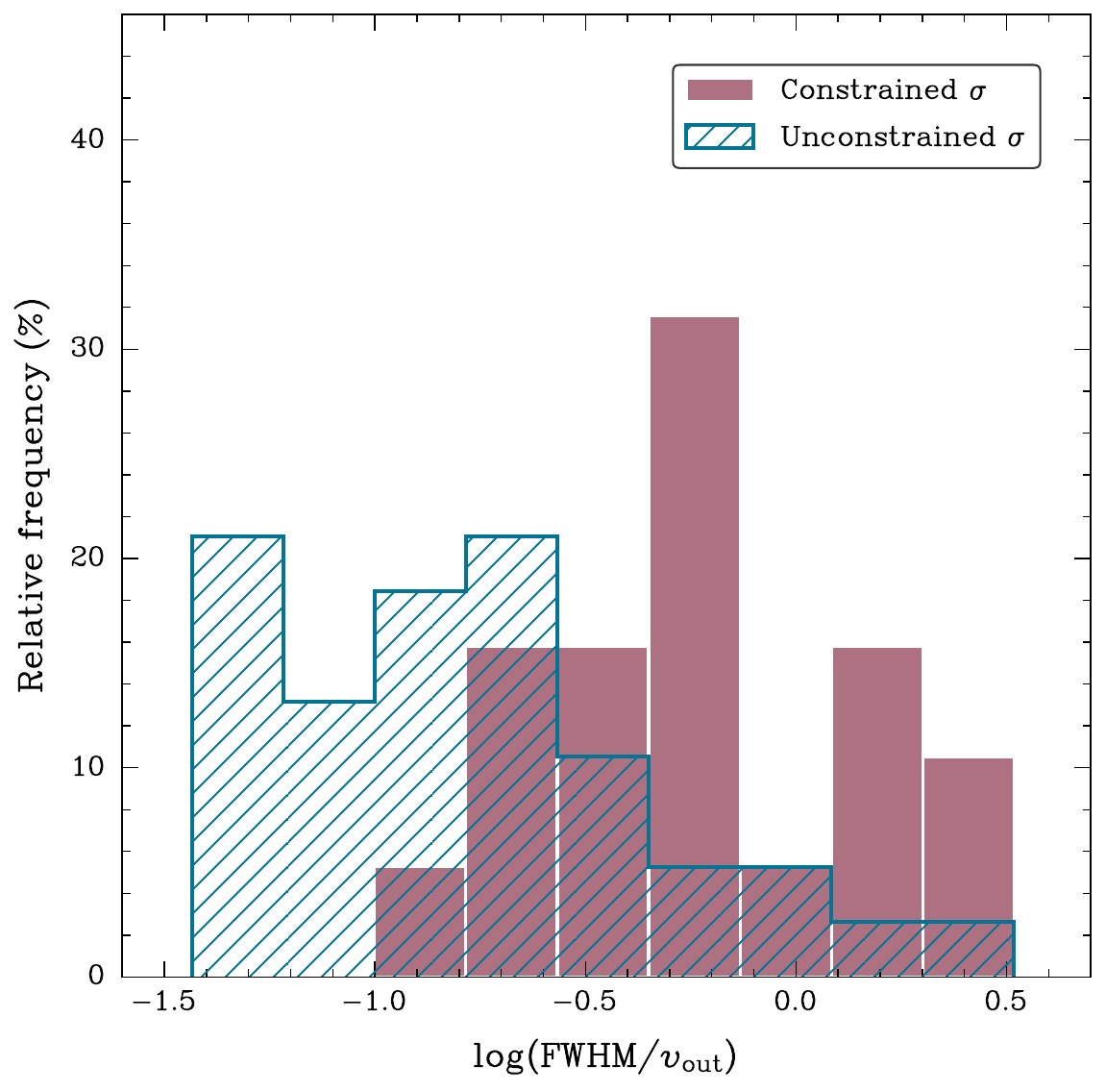}
    \caption{Relative frequency histogram showing the distribution of the ratio $\mathrm{FWHM}/\upsilon_\mathrm{out}$ for the two subsamples with constrained (in red) and unconstrained line width (in blue), respectively. }
    \label{fig:FWHM_ratio}
\end{figure}

Figure \ref{fig:FWHM_ratio} shows the relative frequency histogram of FWHM/$\upsilon_\mathrm{out}$ for both subsamples with constrained and unconstrained values of $\sigma$. The latter is based on the nominal values of the upper limits in FWHM, which, in turn, reflect the upper limits in $\sigma$. In this case, the two distributions appear quite different in shape and spread. To quantify this statement, we adopt both the KS and Mann-Whitney U tests, which return null probabilities comprised between $\sim 10^{-4}$ and $\sim10^{-5}$. This implies that the two subgroups of UFOs in Fig.~\ref{fig:FWHM_ratio} might be characterized by gas outflows with different geometry and/or kinematics.

\subsection{UFO properties across AGN classes}\label{sec:AGNclasses}

The optical classification of our sample sources, obtained from the literature and detailed in Table \TableAone, encompasses a diverse array of astronomical categories. Our primary focus lies in selecting objects identified as QSOs, narrow-line Seyfert galaxies (NLSy1s), and Seyferts. It is notable that the Seyfert category includes both Type 1 and Type 2 classifications, along with various intermediate classes, and also integrates broad line radio galaxies (BLRGs), which constitute approximately $\sim14\%$ of the entire sample.

The inclusion of these categories is substantiated by the observation that Seyfert galaxies, regardless of their specific type, exhibit similar nuclear characteristics, thus demonstrating that the unification model is still mostly valid \citep[e.g.][]{Thean2001, Singh2011, Spinoglio2022}. Furthermore, the optical and X-ray spectral properties of BLRGs show noteworthy parallels with those of Seyfert galaxies, particularly Seyfert 1 types \citep[e.g.][]{Osterbrock1978, Ballantyne2014, Gupta2018}. Finally, recent studies suggest that UFOs powered by RQ or RL sources share similar properties \citep[e.g.][]{Mestici2024}.

\begin{figure*}[ht]
    \centering
    \includegraphics[scale=.325]{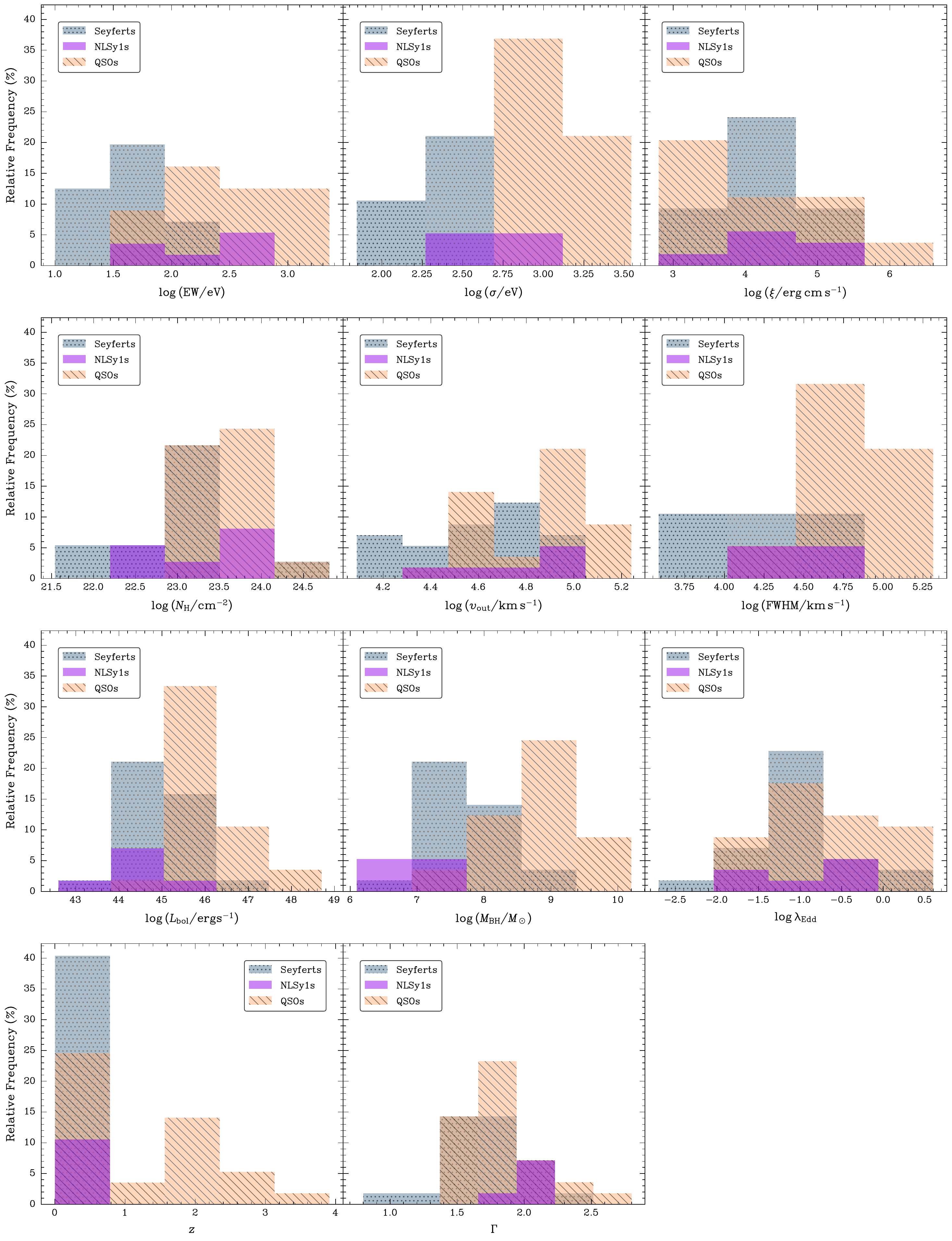}
    \caption{Relative frequency histograms showing the distribution of the phenomenological and physical UFO properties, main AGN parameters and X-ray power-law continuum spectral index across different AGN classes.} 
    \label{fig:Histogram_Classes}
\end{figure*}

Figure \ref{fig:Histogram_Classes} illustrates the distribution of the principal wind phenomenological parameters (centroid energy, $\sigma$ and EW of the Gaussian absorption line) and physical parameters ($\upsilon_\mathrm{out}$, $\log\xi$, and $\log{N_\mathrm{H}}$ from XSTAR) across the various AGN classes outlined above. Upon initial examination, it appears that there might be a gradual transition between Seyferts and QSOs concerning the UFO parameters, with NLSy1s potentially acting as a bridge between these two groups. To rigorously assess this observed effect, we apply both the KS and Mann-Whitney U tests to investigate any possible intrinsic differences between the observed characteristics of UFOs driven by either Seyferts or QSOs. The findings, which are also presented in the same figure, reveal that all UFO parameters, with the exception of $\log{N_\mathrm{H}}$ and $\log{\xi}$, demonstrate significant differences between Seyferts and QSOs. This allows us to decisively reject the null hypothesis that they originate from the same underlying distribution, with a confidence level always exceeding $99.99\%$.

The same figure also shows the distributions of the main AGN parameters and the photon index of the X-ray power-law continuum in the different AGN classes. A significant distinction ($>4\sigma$ c.l.) is identified between Seyferts and QSOs concerning $L_\mathrm{bol}$ and $M_\mathrm{BH}$, aligning with expectations. However, $\lambda_\mathrm{Edd}$ and $\Gamma$ show no such disparity, indicating similar accretion flows and X-ray continuum properties. We highlight that the $\Gamma$ values drawn from the various works are model dependent and have been estimated according to heterogeneous analyses. Therefore, the observed distributions are merely indicative of the general trend, but a homogeneous study is mandatory to detect putative differences among the AGN classes.
The observed redshift distribution pattern results from observational biases (see Sect.~\ref{sec:discussion}).

The outcomes from both the KS and Mann-Whitney U tests, which we used to compare the distributions of Seyferts and QSOs for the quantities in Fig.~\ref{fig:Histogram_Classes}, are presented in Table \ref{tab:distros}.

\begin{table}[t]
    \centering
    \renewcommand{\arraystretch}{1.5}
    \caption{Results of the KS and Mann-Whitney U tests used to compare Seyfert galaxies and QSOs regarding the distributions of their main UFO and AGN parameters, and X-ray continuum. }\label{tab:distros}
    \begin{adjustbox}{max width=\textwidth}
    \begin{tabular}{c c c}
        \multicolumn{3}{c}{Seyferts vs QSOs}                  \\
        \hline
                  &  \multicolumn{2}{c}{$p-\mathrm{value}$}   \\
        \hline
        Parameter & KS test & U test   \\
        (1)       & (2)     & (3)      \\
        \hline\hline
        EW (eV)                                             & $3.2 \times 10^{-6}$  & $3.9 \times 10^{-7}$  \\
        $\sigma$ (eV)                                       & $1.6 \times 10^{-4}$  & $1.6 \times 10^{-4}$  \\
        $\log{(\xi / \mathrm{erg\,cm\,s}^{-1})}$            & $0.13$                & $0.19$                \\
        $\log{(N_\mathrm{H} / \mathrm{cm}^{-2})}$           & $0.043$               & $0.031$               \\
        $\upsilon_\mathrm{out}$ (km s$^{-1}$)               & $3.8 \times 10^{-3}$  & $2.5 \times 10^{-3}$  \\
        FWHM (km s$^{-1}$)                                  & $6.0 \times 10^{-3}$  & $1.9 \times 10^{-3}$  \\
        $\log{(L_\mathrm{bol} / \mathrm{erg\,s}^{-1})}$     & $1.1 \times 10^{-6}$  & $9.6 \times 10^{-7}$  \\
        $\log{(M_\mathrm{BH}/M_\odot)}$                     & $2.8 \times 10^{-6}$  & $3.5 \times 10^{-6}$  \\
        $\log{\lambda_\mathrm{Edd}}$                        & $0.24$                & $0.12$                \\
        $z$                                                 & $1.3 \times 10^{-11}$ & $2.3 \times 10^{-9}$  \\
        $\Gamma$                                            & $0.60$                & $0.25$                \\
        
        \hline
    \end{tabular}
    \end{adjustbox}
\end{table}

\section{Discussion}\label{sec:discussion}

Systematic studies of UFOs line profiles have been hindered by the limited instrumental resolution of the currently operating CCD cameras onboard X-ray telescopes. First, this poses a real challenge in discerning the underlying acceleration mechanism. Indeed, the two main popular models -- namely, MHD and line driving -- would leave a unique imprint on the spectral shape of the absorption line on average: MHD-driven winds typically produce asymmetric profiles with extended blue wings, while line-driven winds can generate red wings \citep[e.g.][]{Fukumura2022, Gandhi2022, Chakravorty2023, Datta2024, Pounds2025}. We note, however, that under certain wind kinematics, line-driven and MHD-driven outflows may produce similar profiles \citep[e.g.][]{Giustini2012}, and recent hydrodynamic simulations of disk winds also predict extended blue wings in some cases \citep[e.g.][]{Waters2021}. 
Despite instrumental limitations, traditional modeling in terms of Gaussian profiles offers a simple but effective solution in the context of sample studies to investigate relationships between the wind spectral signatures. In this work, for example, we find positive correlations, all of them at the $\geq98\,\%$ confidence level, between $\sigma$, EW, and $\upsilon_\mathrm{out}$. These results provide tentative indications that the broadest and/or most intense spectral features correspond to the fastest winds. We note that these indications are not conclusive because they reflect the results gathered by current CCD cameras and will likely be modified by the advent of higher-quality spectroscopy \citep[see, e.g.][]{PDS456}. We also highlight that \--- due to instrumental limitations \--- the sample presented in this paper is naturally affected by a Malmquist bias (see Fig.~\ref{fig:L_z}), implying that at higher $z$ we preferentially observe more luminous AGNs hosting more massive SMBHs, whose environment could differ from that of lower-$z$ sources.  Nonetheless, it is interesting how these results fit into the context of the AGN feedback cycle envisaged by cold chaotic accretion (CCA; e.g., \citealt{Gaspari2013, Gaspari2019, Voit2018, Wittor2020, Wittor2023}). According to the CCA paradigm, UFOs can produce turbulent eddies, which, in turn, are responsible for the condensation of warm filaments and cold gas clouds via nonlinear thermal instability. The condensed gas then rapidly cools and rains onto the central SMBH. In response, the AGN feedback is triggered when the binding energy of the infalling and inelastically colliding clouds is converted into mechanical energy, which is predominantly released into the ambient medium via collimated jets and UFOs. As a result, stronger UFOs are expected to induce stronger turbulence both within and outside the flow \citep[e.g.][]{Wittor2020, Wittor2023}, due to increased interactions with the surrounding medium. Moreover, in the central AGN regions, enstrophy \--- a proxy of solenoidal turbulence, and hence line broadening \--- is even more strongly boosted due to enhanced baroclinic motions (see \citealt{Wittor2023}). Furthermore, CCA is believed to operate self-similarly over different environments \citep[e.g.][]{Gaspari2019}, i.e., independently of $z$, recurrently boosting UFOs over a wide range of parameters.

Our findings in Sect.~\ref{sec:ufo_properties} can also possibly be linked to a clumpy wind scenario \citep[e.g.][]{Takeuchi2014, Dannen2020, Waters2017, Waters2021}. Indeed, the formation of clumpy outflows in AGNs is fundamentally linked to thermal instability (TI), which represents a critical mechanism connecting thermal driving and mass outflow launching. For instance, \citet{Waters2021} demonstrated that radial wind solutions of X-ray-heated flows become inherently prone to developing clumpy structures due to this interconnectedness. The development of clumpy structures occurs through a specific sequence of physical processes involving the formation of what are termed irradiated atmospheric fragments (IAFs). Initially, smooth steady-state solutions become disrupted when buoyancy effects interfere with the stratified structure of the wind. Instead of maintaining a smooth transition layer between the highly ionized disk wind and the cold atmospheric phase below, hot bubbles formed from thermal instability begin to rise and fragment the atmosphere. These bubbles first manifest within large-scale vortices that develop below the transition layer, resulting in the episodic production of distinctive cold phase structures. Upon interaction with the wind, these IAFs advect outward and develop extended crests, with their subsequent disintegration occurring within turbulent wakes that reach high elevations above the disk. This process fundamentally alters the wind structure from a smooth, laminar flow to a complex multiphase medium characterized by significant spatial and temporal variations.
The fragmentation of the wind into clumps naturally increases the velocity dispersion within the outflow. This turbulence broadens the absorption lines, leading to larger line widths. As clumps are advected outward and mixed into the wind, the column density -- and thus, the EW -- along our line of sight can increase, especially at higher velocities where the wind is more turbulent and clumpy. 
Taken at face value, the positive trends involving $\upsilon_\mathrm{out}$, EW, and $\sigma$ are consistent with a clumpy, multi-component absorber in which an increased covering factor and velocity dispersion deepen and broaden the line. Physically, the co-variation can be understood as follows. In the CCA picture, cold clouds condense out of a turbulent hot halo and rain toward the SMBH. Rapid accretion episodes near the ISCO convert part of that inflow energy into UFOs. These UFOs interact with the ambient multiphase medium, entraining gas and driving turbulence. Turbulence and pressure fluctuations, in turn, promote further condensation and fragmentation into clumps. The result is a self-regulated inflow–outflow loop in which feeding (CCA rain) and feedback (UFOs and entrainment) are causally linked and jointly produce clumpy, multi-component outflows whose kinematic and spectral features naturally co-vary. This is non-exclusive: TI/CCA offers one very plausible route to clump formation and turbulence (as supported by a wide literature; see \citealt{Gaspari2020} review), while MHD and line-driven winds may yield complementary line shapes under realistic kinematics.

\subsection{Line Broadening}
Line broadening provides useful insight into the velocity gradient, but this information is often neglected when studying the possible correlations involving the outflow velocity. For example, in their recent work, \citet{yamada2024} assembled a sample of 132 AGNs at $z\lesssim4$ for which at least one blueshifted absorption feature in any of their available X-ray spectra has been reported in the literature as of the end of 2023. This AGN sample, named X-WING, included a total of 583 X-ray winds, incorporating both WAs and UFOs. Considering the distribution of the X-WING winds in the $\upsilon_\mathrm{out}-\log\xi$ plane, the authors claimed the existence of a possibly genuine gap between lowly ionized UFOs and WAs, interpreting this fact as the manifestation of two distinct underlying physical mechanisms driving these outflows. However, this result did not take into account the actual width of the line profile and, thus, the velocity dispersion of the winds. 

In the present work, we find that UFOs can attain a wide range of possible speeds quantified by the FWHM of the Gaussian line, as measured in the rest frame of the wind. The median value of the FWHM for our sources is as large as $\sim40000$ km s$^{-1}$, while the typical uncertainty in $\upsilon_\mathrm{out}$ spans an interval of $\sim7000$ km s$^{-1}$ around the best-fit value. This result highlights the pivotal importance of accounting for line width when addressing relations with $\upsilon_\mathrm{out}$.

However, it may be challenging to decipher the true origin of such a large observed line broadening. One possibility is that the velocity gradient is due to a wind that is launched at different radii or is clumpy. For example, this would naturally arise in either the TI or CCA framework, as discussed above.

Alternatively, according to \citet{Fukumura2019}, the observed line broadening cannot be entirely attributed to the effect of turbulence, but rather it is a manifestation of the wind kinematic field. The authors suggest that at least part of the broadening may reflect a transverse velocity gradient due to Doppler broadening around a putative compact X-ray corona in the proximity of a black hole. In the MHD framework, the magnetic field anchored to the Keplerian disk is naturally capable of imprinting a considerable azimuthal velocity component on the wind. However, this effect can also be in place for radiation-driven winds, assuming that the ejected gas has a nonzero angular momentum or is initially corotating with the disk.

\subsection{UFO location}
Another downside of the resolution of present-day CCDs is the difficulty in constraining the location of the UFO with sufficient precision. To date, such an achievement can only be accomplished using advanced and sophisticated X-ray wind modeling tools like WINE \citep[e.g.][]{Laurenti2021}. However, lower limits $r_\mathrm{min}$ on the launching radius of the UFO can be estimated from simple assumptions, that is, equating the observed $\upsilon_\mathrm{out}$ to the escape velocity \citep[e.g.][]{Tombesi2013, Gofford2015, Chartas2021}. In Sect.~\ref{sec:radius} we considered two groups of AGNs from our sample, which were characterized either by a well-resolved absorption line or by a lack thereof, and calculated the values of $r_\mathrm{min}$ for both groups. 
Interestingly, the $r_\mathrm{min}$ distributions of the two distinct subsamples are quite homogeneous \--- indicating a common spatial origin for the acceleration of such powerful winds \--- and their lower bound lies around $6\,r_\mathrm{g}$, corresponding to the ISCO around a weakly or non-rotating Schwarzschild black hole, showing a truncation toward smaller radii. This is in agreement with the findings in C21 (see Fig.~4 in their paper), who analyzed a sample of 14 moderate-to-high-$z$ lensed QSOs. Apparently, this result still holds when we complement their sample with other independent collections of UFOs from different studies, implying that the observed truncation occurs for a widely heterogeneous AGN population and raising a putative missing-radii problem. Clearly, we must be aware that we observe the projected velocity component of the wind \--- as pointed out by C21 \--- depending on the inclination of the line of sight. Therefore, the true UFO velocity can be larger, leading to smaller $r_\mathrm{min}$ values. 
Although this is true, the putative truncation in Fig.~\ref{fig:rmin} can possibly be explained by the presence of the X-ray corona at smaller radii. Alternatively, the spectral signatures of the fastest winds may be sufficiently broad to become indistinguishable from the underlying continuum.

\subsection{The Seyferts-QSOs dichotomy in UFOs}
In Sect.~\ref{sec:AGNclasses}, intrinsic distinctions between phenomenological and physical UFO properties in Seyfert galaxies and QSOs were evaluated. Significant differences in the distributions of EW, $\sigma$, $\upsilon_\mathrm{out}$, and FWHM were observed, suggesting larger values in QSOs than in Seyferts, with NLSy1s bridging these groups (see Fig.~\ref{fig:Histogram_Classes}). Moreover, we did not detect any clear distinction between these classes in terms of column density and ionization parameter, as is expected from the simple consideration that the outflows considered in this work are all characterized by absorption features originating from similar ionization states.
The different observed behaviors in UFOs cannot be attributed to the properties of the ionizing radiation \--- namely, the X-ray power-law continuum \--- since the distribution of $\Gamma$ values between Seyferts and QSOs is broadly overlapping, evidence that is also supported by previous findings \citep[e.g.][]{Dai2004, Piconcelli2005, Just2007, Dadina2008, Huang2020, Ojha2020}.
At the same time, the discrepancies are hardly caused by intrinsic differences in the accretion rate. In fact, if on the one hand the large uncertainty in $\lambda_\mathrm{Edd}$ makes it difficult to detect distinct patterns between different AGN classes, there is a growing consensus that the slope of the ionizing X-ray power-law continuum is weakly affected by variations in $\lambda_\mathrm{Edd}$ \citep[e.g.][]{ai2011, liu2016, martocchia2017, Trakhtenbrot2017, Tortosa2018, Laurenti2022, kamraj2022, Trefoloni2023, Laurenti2024, Serafinelli2024}.
The origin of the observed different UFO properties in Seyferts and QSOs would then be explored through the intrinsically different properties characterizing these classes -- namely, $M_\mathrm{BH}$ and especially $L_\mathrm{bol}$. 

One intuitive way to reconcile the distinctive trends found in Seyferts and QSOs is through the TI framework. In this picture, systems with higher luminosities, such as QSOs, produce a more intense radiation field that creates more prominent zones where TI can operate, leading to more extensive clump formation and greater turbulence levels. This, in turn, can produce broader absorption lines (larger $\sigma$ and FWHM) and increased opacity effects (larger EW). It is crucial to distinguish the mechanism shaping the wind's structure from the one providing its bulk acceleration. We refrain from attributing the large $\upsilon_\mathrm{out}$ solely to thermal pressure. Instead, we propose that processes like TI/CCA establish the multiphase, clumpy structure affecting EW and $\sigma$, while a more powerful mechanism, such as radiation or MHD driving, plausibly provides the bulk acceleration to mildly relativistic speeds.

Alternatively, the CCA model can offer a complementary explanation. In this scenario, bolometric luminosity scales with the central mass accretion rate ($L_\mathrm{bol} \propto \dot{M}_\mathrm{acc}$), which is tightly linked to the mass supplied from larger scales ($\dot{M}_\mathrm{supply}$). A higher supply rate in QSOs leads to a denser ``rain'' of condensing gas filaments. During these high-supply phases, more frequent cloud-disk interactions increase turbulent broadening and the covering factor of the absorbing gas, naturally elevating FWHM and EW. Supersonic collisions between these clumps would generate shocks, which can locally heat the gas. However, our ensemble analysis shows no significant class-level shifts in $\log\xi$ or $\log{N_\mathrm{H}}$ between Seyferts and QSOs. This indicates that while localized shocks can raise ionization, the Fe XXV/XXVI phase remains comparable across classes.

These empirical relationships align with the characteristics of clumpy, turbulent winds in TI/CCA-like settings. However, they may also be compatible with MHD and line-driven disk winds. For example, radiation-hydrodynamic simulations of line-driven disk winds predict that the outflow energetics scales with bolometric luminosity \citep[e.g.][]{Nomura2017}, offering a different framework in which distinct UFO properties might naturally arise as a consequence of luminosity-dependent differences among AGN classes. Our current analysis does not discriminate among these scenarios.

\subsection{XRISM simulations}

\renewcommand*{\ttdefault}{qcr}

\begin{figure}[t]
    \centering
    \includegraphics[width=\columnwidth]{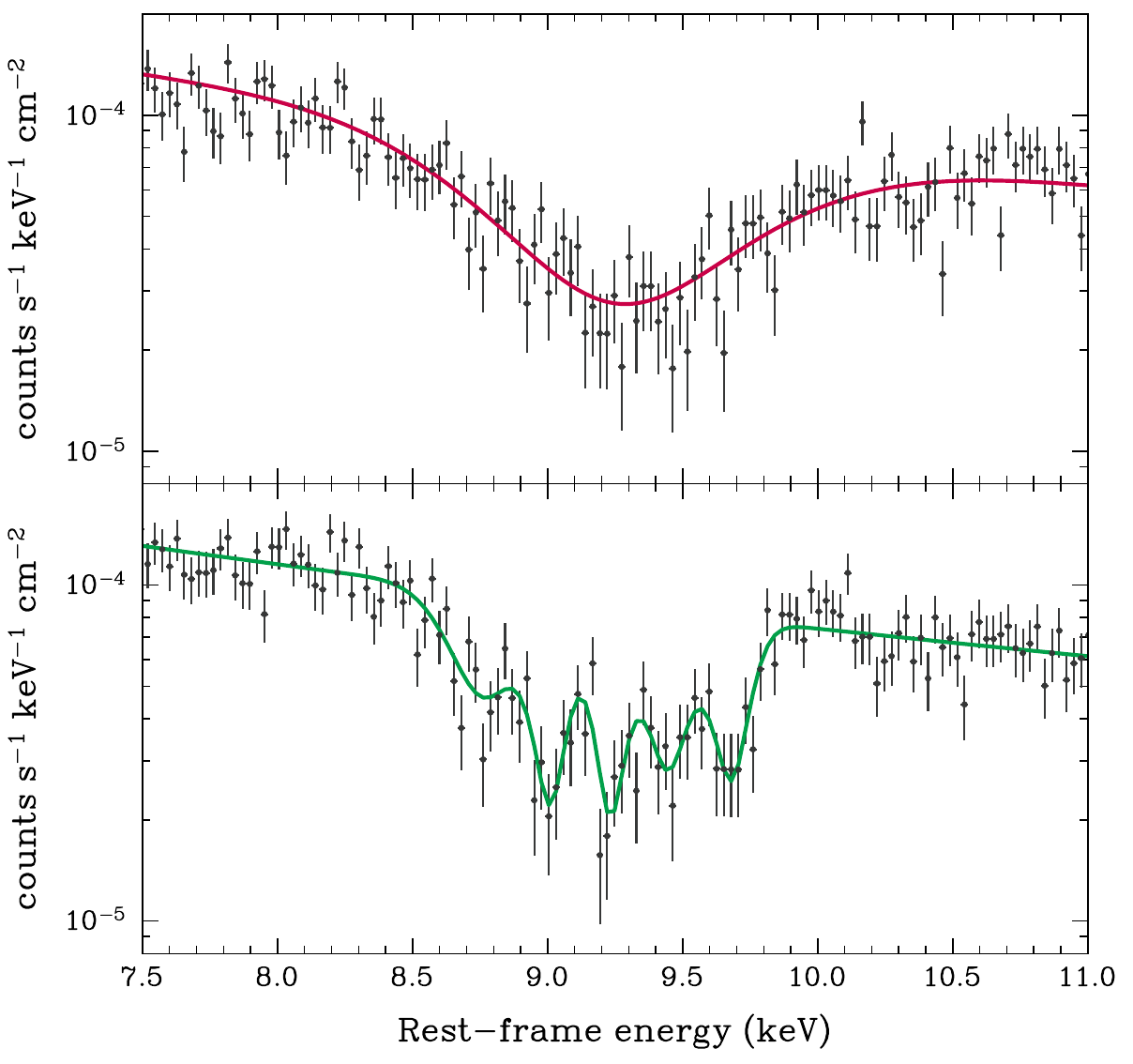}
    \caption{Results of two simulations of a 100 ks \emph{XRISM} observation zoomed in on the blueshifted Fe K absorption signature. The X-ray continuum consists of a power law modified by Galactic absorption. \emph{Top panel}: the UFO absorption feature is modeled with a single broad absorption Gaussian line. \emph{Bottom panel}: the broad absorption line is superseded by five narrow Gaussian absorption lines to test the capabilities of \emph{XRISM}'s Resolve calorimeter to detect structures within the outflow profile.}
    \label{fig:xrism_simulations}
\end{figure}

Finally, we must be aware that an apparently broad and smooth line profile might be observed as the result of the combination of a forest of narrower absorption features, as recently discovered for the first time by the X-Ray Imaging and Spectroscopy Mission (\emph{XRISM}; \citealt{xrism2020}) for the UFO in the quasar PDS 456 \citep[][]{PDS456}. For PDS 456, the high-resolution spectrometer Resolve onboard \emph{XRISM} enabled the discovery of five discrete velocity components outflowing at $20{-}30\%$ of the speed of light, supporting models envisaging an inhomogeneous UFO consisting of up to a million clumps. In line with these results, we examined \emph{XRISM}'s capability to replicate a comparable scenario in the AGN sample discussed in this paper, benefiting from its innovative instrumentation. 
Specifically, the Resolve spectrometer can reach an exceptional resolution of $\sim5$ eV over the whole effective\footnote{The \emph{XRISM} Resolve instrument's Gate Valve (GV) has not yet opened, thereby shifting Resolve's energy band from $E=0.3{-}12$ keV to $E=1.7{-}12$ keV and lowering the effective area. This has a negligible effect on the energy interval corresponding to Fe K transitions.} bandpass ($E=1.7{-}12$ keV), thanks to its X-ray calorimeter. We downloaded the latest publicly available canned response files and used XSPEC (\texttt{v12.14.1}; \citealt{Arnaud1996}) to simulate a 100 ks \emph{XRISM} observation, assuming a flux of $F_\mathrm{2-10\,keV}\sim 5 \times 10^{-12}$ erg s$^{-1}$ cm$^{-2}$, corresponding to the median flux level of our AGN sample.
We first simulated a spectrum characterized by a simple power-law continuum modified by Galactic absorption, plus a broad absorption line consisting of a Gaussian whose centroid, line width, and EW represent the median values of the corresponding quantities calculated for the 19 AGNs in our sample with a well-constrained absorption line. The median values of redshift and $N_\mathrm{H,gal}$ for the same sources are used as well. The baseline model is described by \texttt{TBabs\,$\cdot$\,(zpowerlw + zgauss)} in XSPEC, and the result is shown in the top panel of Figure \ref{fig:xrism_simulations}.
Then we considered the case in which the broad absorption line is superseded by five equally spaced narrow absorption lines whose cumulative width and EW match those of the original broad line. Interestingly, \emph{XRISM} allows us to distinguish the five absorption components and constrain their widths and EWs with relative errors as small as ${\sim}\,20{-}30\%$ (see bottom panel of Figure \ref{fig:xrism_simulations}). Modeling these five narrow absorption lines with a single broad absorption feature would lead to a worse fit, quantified by a variation of $\Delta{\chi^2}=77$ for a corresponding change of $\Delta\nu=12$ in degrees of freedom.

\section{Conclusions}\label{sec:conclusions}

We collected a sample of 122 UFO detections, comprising a total of 57 AGNs spread across wide intervals in redshift ($z\lesssim4$), bolometric luminosity ($10^{43} \lesssim L_\mathrm{bol} \lesssim 10^{49}$ erg s$^{-1}$), black hole mass ($10^6 \lesssim M_\mathrm{BH}/M_\odot \lesssim 10^{10}$), and Eddington ratio ($-2.7 \lesssim \log{\lambda_\mathrm{Edd}} \lesssim 0.6$). 
We combined the results from both the phenomenological and physical modeling of the UFO absorption lines and found the following results.
\begin{itemize}
    \item A positive correlation is emerging between the Gaussian line width, EW, and the outflow velocity calculated from photoionization modeling with XSTAR. Specifically, we get a tentative indication that the broadest spectral features are characterized by the largest EW and are associated with the fastest winds. These results reflect the capabilities of current CCD cameras and will likely be, at least partially, updated with the advent of higher-quality spectroscopy. 
    
    \item The FWHM of the Gaussian line, once estimated in the rest frame of the wind and converted into velocity space, is indicative of the UFO velocity dispersion. This dispersion is typically much larger than the statistical uncertainty in $v_{\rm out}$. Therefore, when studying correlations involving the outflow velocity, it is important to account not only for the centroid velocity but also for the velocity dispersion, since objects with similar $v_{\rm out}$ may differ substantially in their kinematics. Moreover, a larger velocity width must be considered when estimating the energetics of the UFOs, as key quantities such as the mass outflow rate, momentum rate, and kinetic power all scale with velocity. In this sense, integrating these quantities over the velocity distribution provides a more realistic estimate than relying solely on the centroid velocity. 
    
    \item The properties of UFOs, on both phenomenological and physical grounds, such as EW, $\sigma$, FWHM, and $\upsilon_\mathrm{out}$, appear to be distinct among Seyfert galaxies and QSOs, with NLSy1s bridging these two groups. Apparently, this is likely attributable to their intrinsic differences in terms of $M_\mathrm{BH}$ and $L_\mathrm{bol}$. These empirical relations are consistent with clumpy, turbulent winds in TI/CCA-like environments, but they may also be compatible with MHD and line-driven disk winds. 
    
    \item Simulations show that exceptionally high-resolution spectroscopy with \emph{XRISM} (or the future \emph{NewAthena} mission) will allow us to shed definitive light on the properties of the UFO spectral features while hopefully providing insights into the wind energetics, geometry, and kinematics as well.
        
\end{itemize}

\section*{Data availability}
\phantomsection
\label{sec:data_availability}
\renewcommand*{\ttdefault}{cmr}
Tables \TableAone and \TableBone are only available in electronic form at the CDS via anonymous ftp to \href{http://cdsarc.u-strasbg.fr/}{\nolinkurl{cdsarc.u-strasbg.fr}} (130.79.128.5) or via \url{http://cdsweb.u-strasbg.fr/cgi-bin/qcat?J/A+A/}.

\begin{acknowledgements}  
    We thank the anonymous referee, whose constructive comments helped us improve this paper.
    This work is based on observations obtained with \emph{XMM-Newton}, an ESA science mission with instruments and contributions directly funded by ESA member states and the USA (NASA), \emph{Suzaku}, a collaborative mission between the space agencies of Japan (JAXA) and the USA (NASA), \emph{NuSTAR}, a project led by the California Institute of Technology, managed by the Jet Propulsion Laboratory and funded by the National Aeronautics and Space Administration, and \emph{Chandra} X-ray observatory. We acknowledge funding from the European Union - Next Generation EU, PRIN/MUR 2022 (2022K9N5B4). MG acknowledges support from the ERC Consolidator Grant \textit{BlackHoleWeather} (101086804). RS acknowledges funding from the CAS-ANID grant number CAS220016.
\end{acknowledgements}

\bibliographystyle{aa} 
\bibliography{biblio}

\end{document}